\documentstyle[12pt,aaspp4]{article}

\def\etal{{et al.}}
\def\asca{{\it ASCA}}

\newbox\grsign \setbox\grsign=\hbox{$>$} 
\newdimen\grdimen \grdimen=\ht\grsign
\newbox\simlessbox \newbox\simgreatbox \newbox\simpropbox
\setbox\simgreatbox=\hbox{\raise.5ex\hbox{$>$}\llap
     {\lower.5ex\hbox{$\sim$}}}\ht1=\grdimen\dp1=0pt
\setbox\simlessbox=\hbox{\raise.5ex\hbox{$<$}\llap
     {\lower.5ex\hbox{$\sim$}}}\ht2=\grdimen\dp2=0pt
\setbox\simpropbox=\hbox{\raise.5ex\hbox{$\propto$}\llap
     {\lower.5ex\hbox{$\sim$}}}\ht2=\grdimen\dp2=0pt

\begin{document}

\title{X-ray Signatures of an Ionized Reprocessor in the 
Seyfert galaxy Ton S~180} 

\author {T.J.Turner \altaffilmark{1,2}, 
I.M. George \altaffilmark{1, 2}, 
K. Nandra \altaffilmark{1, 3} }

\altaffiltext{1}{Laboratory for High Energy Astrophysics, Code 660,
	NASA/Goddard Space Flight Center,
  	Greenbelt, MD 20771}
\altaffiltext{2}{Universities Space Research Association}
\altaffiltext{3}{NAS/NRC Research Associate}

\slugcomment{To be submitted to {\em The Astrophysical Journal}}

\begin{abstract}

We discuss the hard X-ray properties of the Seyfert galaxy 
Ton S~180, based upon the analysis of \asca\ data. 
We find the X-ray flux varied by a factor $\sim 2$ 
on a time scale of a few thousand seconds. The source showed 
significantly higher amplitude of variability in the 0.5--2 keV 
band than in the 2--10 keV band. The continuum is adequately 
parameterized as a $\Gamma \sim 2.5$ 
power-law across the 0.6--10 keV band. 
We confirm the recent discovery of an emission line 
of high equivalent width, due to Fe K-shell emission 
from highly-ionized material. These {\it ASCA} data show the Fe line 
profile to be broad and asymmetric and tentatively suggest it is 
stronger during the X-ray flares, consistent with 
an origin from the inner parts of an accretion disk. 
The X-ray spectrum is complex below 2 keV, possibly due to 
emission from a blend of soft X-ray lines, which would support the 
existence of an ionized reprocessor, most likely due to 
a relatively high accretion rate in this source.

\end{abstract}

\keywords{galaxies:active -- galaxies:nuclei -- X-rays: galaxies -- 
galaxies: individual(Ton S 180) }

\section{Introduction}

\label{sec:intro} 
Ton~S~180  (PHL 912) 
is the brightest Active Galactic Nucleus (AGN) detected by 
the Extreme Ultraviolet Explorer ({\it EUVE} ; 
Bowyer \etal\ 1996) as it is a strong emitter in the 
ultraviolet regime and also has a 
low Galactic column density along the line-of-sight 
($N_H=1.52 \times 10^{20} {\rm cm}^{-2}$; Stark \etal\ 1992). 
This source has 
FWHM H$\alpha$ and H$\beta \sim 900 {\rm km\ s}^{-1}$ and a redshift
z=0.06198 (Wisotzki \etal\ 1995). The absolute 
magnitude is M$_{\rm B}=-23.1$, on the borderline 
often used to (arbitrarily) distinguish Seyfert galaxies and quasars. 

A terminology has developed which names a subclass of  
Seyfert 1 galaxies as  narrow-line Seyfert 1 galaxies (NLSy1). This name is 
somewhat confusing: objects classed as NLSy1 are loosely defined as having 
the narrowest optical permitted lines in the distribution covered by Seyfert 1
galaxies. 
Typically the widths of the broad components of  H$\alpha$ and H$\beta$ are 
only slightly broader 
than the forbidden lines, FWHM H$\beta <2000$ km/s (although 
some authors consider only sources with FWHM of the Balmer lines 
$< 1500$ km/s to be NLSy1) and 
\verb+[+O{\sc iii}\verb+]+/H$\beta\ < 3$, to distinguish them from
Seyfert~2 galaxies.  Strong Fe{\sc ii} emission is often 
present in sources with the narrowest permitted lines. 
An alternative, less common name for these objects is I ZW 1 objects,
which is a well-studied example of this type (Halpern \& Oke 1987). 
However, the emerging consensus of opinion seems to be 
that there is a distribution of widths of the permitted 
lines, and other
properties, and so the definition of a subclass can be considered 
somewhat arbitrary (Lawrence \etal\ 1997).  Nevertheless, the properties 
of Ton S~180 have lead it to be commonly referred to as a NLSy1.

The strong anti-correlation between soft X-ray slope and FWHM
H$\beta$ (Boller \etal\ 1996) is of interest and may explain 
why the list of bright AGN detected by {\it EUVE} is dominated by sources 
with relatively narrow permitted lines. 
Brandt \etal\ (1997) showed the correlation also exists between 
 hard X-ray spectrum (2--10 keV slope) and FWHM H$\beta$. It has been 
speculated that these effects may be due to a difference 
in a fundamental parameter such as the accretion rate (e.g. Pounds et al
1995). Large 
amplitude and rapid X-ray variability also appears most common 
in objects with narrow permitted lines, and could 
extend into the extreme ultraviolet regime.  
For example, Ton S~180 
has been reported to show a ``doubling time'' of 12 hours in an {\it EUVE} 
observation (Hwang \& Bowyer 1997). 

The {\it ROSAT} PSPC made 15 pointed observations of Ton S~180, and the source 
was also detected in PSPC survey data 
taken during 1990 Dec 16-18 (Fink \etal\ 1997). The source showed flux 
variability up to a factor of $\sim 2$ on timescales of hours to 
years, including some flare-type outbursts. Fink \etal\ find 
a complex spectrum which can be 
parameterized by a power-law $\Gamma=3.1^+_-0.05$ 
plus a black-body of temperature $kT=15.7^+_-2.5$ eV, dominating below 
0.3 keV, both absorbed by a 
column consistent with the Galactic line-of-sight value. However, 
{\it EUVE} yields a flux density $\nu{\rm F}_{\nu}=3.9 \times 10^{-11} 
{\rm erg\ cm}^{-2} {\rm s}^{-1}$ at 0.14 keV (Vennes \etal\ 1995), a factor 
of five lower than the extrapolation of the model to the PSPC data
(Fink \etal\ 1997). 
The PSPC data show that the spectrum of the source was consistent with 
being constant across many of the observations, although a flattening 
of the power-law component was noted between the observations 
on 1992 Dec 18 and 1993 Jan 10, when 
the source was at approximately the same flux-state. 

Comastri \etal\ (1998) confirmed 
the narrow widths of the optical permitted lines using 
a spectrum taken at the ESO 1.52m telescope and presented the results along 
with the 
{\it BeppoSAX} observation of Ton S~180 taken on 1996 Dec 3. 
The effective exposure time 
was $\sim 25$ks with the Medium Energy Concentrator
Spectrometer (MECS) which covers 1.3 -- 10 keV, and $\sim 12$ ks with the 
Low Energy Concentrator Spectrometer (LECS) which covers 0.1 -- 10 keV.
The {\it BeppoSAX} paper reported the first spectrum of 
Ton S~180 above 2 keV, 
which could be parameterized adequately using a double 
power-law model with $\Gamma\sim 2.3$ above 2 keV, and 
$\Gamma\sim 2.7$ below 2 keV. Comastri \etal\ (1998) found significant iron 
K$\alpha$ emission at a rest-energy $\sim 7$ keV, suggesting the line has an origin in 
ionized material. 

Here we present the results of the {\it ASCA} observation of 
Ton S~180, which comprises the only other hard X-ray information available 
to date.  In \S3 we discuss the time variability of the source. \S4 
details the mean X-ray spectrum and in \S5 we present a tentative 
result on spectral variability in the source. 

\section{ASCA Observations and Data Reduction}

{\it ASCA} (Makishima et al. 1996) has four co-aligned 
X-ray telescopes with two solid-state imaging spectrometers 
(SISs; Burke \etal\ 1994) and two gas imaging spectrometers 
(GISs; Ohashi \etal\ 1996) 
in the focal plane, yielding effective bandpasses 
$\sim$0.4--10~keV and  $\sim$0.8--10~keV respectively. 
Ton S~180 was observed by {\it ASCA} on 1996 July 10 -- 11.
The data were reduced in the same way as the Seyfert galaxies
presented in Nandra \etal\ 1997a (N97a) and Turner \etal\ 1997. For details of
 the data reduction method see N97a. Data screening yielded 
effective exposure times of $\sim 44$ ks in the SIS and 
$\sim 49$ ks in the GIS instruments. SIS response matrices were created 
using \verb+SISRMG v1.1+. 
In this paper, all energy ranges are given in the observers frame, unless 
otherwise noted. 

\section{Time Variability}

The flux in the 0.5 -- 2 keV band 
was $1.1 \times 10^{-11} {\rm erg\ cm}^{-2} {\rm s}^{-1}$
during the {\it ASCA} observation, similar to 
some of the flux states observed by {\it ROSAT} (Fink \etal\ 1997). 
The 2 -- 10 keV flux was 
$6.0 \times 10^{-12} {\rm erg\ cm}^{-2} {\rm s}^{-1}$ 
inferring a luminosity of 
$\sim 10^{44} {\rm erg\ s}^{-1}$ (assuming H$_{\rm 0}=50 
{\rm km\ s^{-1}\ Mpc^{-1}} $, 
q$_{\rm 0}=0.5$). These {\it ASCA} fluxes are determined assuming 
a spectral model consisting of  a power-law plus 
Gaussian ``hump'', as  detailed in \S4.1. We note that any 
model providing a good fit to the data will yield a flux which is 
accurate to $\sim 10\%$. For comparison, the  
{\it BeppoSAX} observation found the source to have a 2 -- 10 keV flux
$4.2 \times 10^{-12} {\rm erg\ cm}^{-2} {\rm s}^{-1}$ (Comastri \etal\ 1998). 

Fig.~1 shows soft (0.5 -- 2 keV) and hard 
(2 -- 10 keV) light curves, constructed using 128 s bins, 
along with their ratio. 
Ton S~180 shows variations in flux by a factor $\sim 2$, with the most 
significant variations evident in the 0.5 -- 2 keV band. 
We investigated the ``excess-variance'' ($\sigma^2_{rms}$) for the
light curves. This quantity is a measure of the amplitudes 
of deviations from the mean flux, in excess of those expected 
from statistical scatter within the light curve (for a 
detailed definition of this quantity see N97a). The 
light curve taken over the full 0.5--10 keV band 
shows  $\sigma^{2}_{rms}(0.5-10 keV)=0.026^+_-0.003$. 
In Fig.~2a we show 
$\sigma^{2}_{rms}$(0.5 -- 10 keV) versus luminosity 
for Ton S~180 compared to the Seyfert~1 sample of N97a; 
three sources for which it could be 
measured in the Seyfert~2 sample (Turner \etal\ 1997) and 
some recent measurements for NGC~3783 and NGC~3227 (George \etal\ 1998a, 
1998b). Ton S~180 clearly 
shows larger $\sigma^{2}_{rms}$ (0.5 - 10 keV) than expected for a 
typical Seyfert~1 galaxy of that luminosity. 

Next we examined $\sigma^{2}_{rms}$ (2 -- 10 keV) i.e. that measured 
in the hard band alone, and the results are shown in  Fig.~2b. 
In this band, Ton S~180 has $\sigma^{2}_{rms}$(2 -- 10 keV)$=0.013^+_-0.005$ 
which 
is not so significantly separated from the rest of the 
distribution, compared to the "full-band" result (Fig.~2a). 
We also note that $\sigma^{2}_{rms}$  is significantly greater 
(at 99\% confidence) in the 0.5 -- 2 keV band 
than in the 2 -- 10 keV band, with 
$\sigma^2_{rms}$(0.5 -- 2 keV)$=0.028^+_-0.003$.
Fig.~3 shows soft versus hard $\sigma^{2}_{rms}$ for 
Ton S~180 and the Seyfert~1 sample from N97a. Ton S~180 is one of 
several sources showing a greater $\sigma^{2}_{rms}$ in the soft band 
than the hard. 

\section{The Spectrum} 

While the difference in soft and hard variability properties means 
spectral variations are occurring, the 
hardness ratio is quite flat (Fig.~1) indicating that any spectral variability is 
of low amplitude and may be difficult to measure via spectral analysis. 
Consequently, we first discuss the mean spectrum from the observation. 

We consider only the data in the 0.4 -- 10 keV band.  
However, SIS data below 0.6 keV were excluded from the spectral
analysis 
as it is commonly accepted that there are uncertainties 
associated with the {\it ASCA} calibration in that band.
Whilst the SIS spectral calibration is 
suspect in this band, we do make use of the fact that 
the error is considered to be $\lesssim 20$\%, and 
that it causes the data to lay systematically low.
Thus these data can sometimes be used to indicate which 
of several models may be most applicable. For example, if data in the 
0.4-0.6 keV band lie above 
the extrapolation of our model, 
then the source spectrum most likely does turn-up in the 
0.4-0.6 keV range. 
Hence we take the approach of indicating 
where these data lie, although we never use them in a fit. This allows the 
reader to draw their own conclusions about the shape of the spectrum 
below 0.6 keV. 

\subsection{Exclusion of the Fe K-shell regime}

In this section we 
excluded the (rest-frame) 5.3-8.0 keV data, to temporarily remove the 
channels in which we expect to observe iron K$\alpha$ emission in this 
source, so we could more easily parameterize the continuum shape. 

Neither single nor double power-law models (attenuated by a 
column of neutral material) provided an acceptable fit due 
primarily to complexity in the $\sim 0.6-1$ keV 
regime. The single power-law yielded 
$\chi^2=953$  for 747 degrees of freedom ($dof$), the double power-law gave 
$\chi^2/dof=905/745$
Fig.~4a shows the ratio of the data to a power-law model with the 
0.4-0.6 keV SIS data overlaid in dotted typeface. 

The  column density of the neutral absorber was consistent with the Galactic 
line-of-sight value 
($N_H=1.52 \times 10^{20} {\rm cm}^{-2}$; Stark \etal\ 1992) in the 
aforementioned fits 
and hence was fixed at that value subsequently. 
As shown in Fig.~4a the largest deviation is suggestive of an emission feature 
peaking just below 1 keV. 
An adequate fit is achieved when a broad gaussian 
emission line is added to the model. This gives an emission line 
rest-energy 
$E=0.82^{+0.20}_{-0.35}$keV,  $\sigma=0.25^{+0.17}_{-0.03}$keV 
and line flux of 
$1.27 \times 10^{-3} {\rm photons\ cm}^{-2} {\rm s}^{-1}$, 
yielding $\chi^2/dof=768/744$, an improvement 
of $\Delta \chi^2=185$ compared to a fit without this component 
(all errors are 90\% confidence for the relevant number of interesting 
parameters in each case).   
Fig.~4b shows the data/model ratio after the fit. The 
improvement is immediately obvious. 
When the gaussian component 
is introduced into the model the underlying continuum is 
$\Gamma=2.50^+_-0.08$. We note that with such a model, we 
find no requirement for a second power-law within the 0.6 -- 10 keV 
band, although there is a suggestion of a spectral softening 
below 0.6 keV (as expected on the basis of {\it ROSAT} and {\it BeppoSAX} 
results). 
The equivalent width (EW) of the line was $155^{+262}_{-65}$ eV, and 
could represent the sum of contributions from species such as ionized Ne 
and Fe-L emission. To parameterize the emitting gas, we fit the 
data using the 
``mekal'' model for emission from a thermal plasma, 
which is based on Mewe \& Kaastra 1992, this includes 
important iron L calculations (Liedahl, Osterheld \& Goldstein 1995) 
and we assumed cosmic abundances. This fit equated the 
soft emission to that expected from an optically thin plasma 
in collisional equilibrium with kT=0.84 keV, 
although the overall fit gave $\chi^2/dof=798/745$ 
(a 9\% probability of achieving this fit statistic by chance); 
inferior to that provided using a simple gaussian parameterization. 
The inferiority of the mekal model could be due to the 
assumption of cosmic abundances which leads to the prediction of observable 
O, Si and S lines from the gas, or may indicate the emission is 
from photoionized gas. These data do not allow us to distinguish between these 
possibilities. 

Alternatively, the soft line could represent a single, 
relativistically-broadened line. 
We fit it using 
the disk-line model profile of Fabian \etal\ (1989). That model 
assumes a Schwarzschild geometry and computes the line profile. 

We assumed the line originates from a disk between 
radii, r, from 6 to 1000 r$_g$, where the gravitational 
radius, $r_g=GM/c^2$ for 
a black hole of mass $M$. The line emissivity 
function was assumed to be r$^{-2.5}$, typical of 
Seyfert~1 galaxies (N97a). The inclination 
of the disk relative to the observer is  
defined such that $i=0$ is a face-on disk. 
The fit was inferior to the gaussian model ($\chi^2/dof=786/744$) yielding 
a rest-energy $E=0.86^+_-0.30$ keV, an inclination $i > 77^{\rm o}$, 
$\Gamma=2.58^+_-0.08$ with equivalent width $EW=88^{+130}_{-33}$ eV. 
The poor constraints on energy do not allow us to unambiguously identify
the line, although we note that O{\sc viii}, Ne{\sc ix}, Ne{\sc x} and 
emission from the Fe L-shell are expected to provide strong emission at
energies consistent with the fitted value. 

We also considered alternative models based upon the possibility that the 
observed shape of the soft X-ray spectrum is due to the presence of several 
strong absorption edges. 
 First we fit the data to a power-law plus 
two  absorption edges. This yielded a rest-energy 
$E_1=0.51^{+0.16}_{-0.01p}$ keV,
where $p$ indicates an error range reached the maximum or minimum 
allowed value for the parameter. The optical depth of this edge was 
$\tau_1=0.44^{+25}_{-0.35}$. The second edge could be 
characterized with a rest-energy $E_2=1.42^{+0.05}_{-0.06}$ keV,
$\tau_2=0.25^{+0.05}_{-0.06}$ and the fit gave $\chi^2/dof=841/743$ 
This was clearly less satisfactory than the gaussian emission model 
(Fig.~4c). 
However, a detailed consideration of this model is interesting as 
three Seyferts with similar characteristics to 
Ton S~180 have shown deep absorption features close 
to 1~keV. Leighly \etal\ (1997) interpreted those as evidence for 
absorption in outflowing material with velocities in the range 
0.2-0.6c. In Ton S~180, 
the lower energy edge is consistent with neutral 
oxygen. However, while there are several edges that could be identified 
with the feature at 1.42 keV, none of them would be expected to arise 
from neutral material. Thus we consider whether the two edges might 
both arise from ionized material, flowing into or out from the 
nucleus. 
However, fits to a warm absorber found no satisfactory solution, even 
when the ionization-state and redshift of the absorber were totally 
unconstrained such that relativistic velocities were allowed as 
possible 
solutions. The failure to find a satisfactory fit, even allowing 
a strong blueshift for the absorber, leads us to conclude that 
if the features are due to absorption edges, then 
they are not consistent with both arising in a single zone of 
outflowing material. 

We also tried a broken power-law model 
for the continuum with a single  absorption edge (c.f. Fig.~4a). 
This model provides an adequate description of the data, 
yielding a fit-statistic $\chi^2/dof=770/743$ 
The fit gave $\Gamma_{soft}=3.66^{+0.28}_{-0.27}$ 
meeting  $\Gamma_{hard}=2.56^{+0.07}_{-0.08}$ at
a break point of $1.50^{+0.09}_{-0.08}$ keV with an edge 
of depth $\tau\sim 1.50^{+0.32}_{-0.48}$ at $E=0.61^{+0.02}_{-0.02}$ keV. 
However, when this model is extrapolated down to 0.4 keV, and 
compared to the SIS data, there is clearly a huge discrepancy. Fig.~4d) 
shows the data/model ratio, which indicates that 
the extrapolation of this model down to 
0.4 keV overpredicts the flux by a factor of $\sim 3$ compared to the data. 
This mis-match far exceeds any {\it ASCA} calibration uncertainty 
and indicates that this is an unsatisfactory model (c.f. Orr \etal\ 1998). 

Fink \etal\ (1997) also noted similar spectral complexity 
in PSPC spectra of Ton S~180, although their interpretation was different, 
based upon the limited spectral range available using the PSPC. 
Their favored model was an absorbed power-law, plus a black body of 
temperature $kT \sim 16$ eV, dominating below 0.3 keV. The latter 
would not be visible in the {\it ASCA} band but in any case 
the {\it ROSAT} model 
of Fink \etal\ (1997) is inconsistent with our {\it ASCA} data. 
Specifically, their 
power-law component is too steep to be consistent with our hard X-ray data.
However, the evidence for spectral features looks similar in both 
instruments, so it is interesting to examine some of 
the PSPC data in light of the {\it ASCA}  
result. We selected the 1993 June 16 data, which was the last and 
longest observation of Ton S~180 by the PSPC. This 
revealed a photon index $\Gamma=2.96^+_-0.03$ with 
no intrinsic absorption over the Galactic line-of-sight value. 
The data/model ratio confirms the spectral complexity 
(Fig.~5), as noted by Fink \etal\ (1997). 
Those authors modeled the shape we see as a juxtaposition of 
two continuum components, but we favor an emission feature. 
Our power-law plus gaussian model from the {\it ASCA} data is 
consistent with the PSPC data. The strength of the  
gaussian component cannot be usefully constrained fitting the latter 
but is consistent with the {\it ASCA} result. 
Thus, these PSPC data increase our confidence as 
to the reality of the feature seen in the SIS data.
The PSPC data also indicate the presence of additional spectral features, 
perhaps deep absorption close to 0.4 keV. 
Although the PSPC has proved a reliable ``finder'' for 
interesting sources, it 
does not allow us to unambiguously determine the origin of
the spectral complexity. We cannot distinguish between 
our solution and the model  
favored by Fink \etal\ (1997), thus we do not attempt to model 
the PSPC data in more detail. 

No soft X-ray feature was detected in the {\it BeppoSAX} observation
(Comastri \etal\ 1998), but the exposure in the LECS was very low. 
Also, the LECS has a slightly lower energy resolution, $FWHM \sim 200$ eV at 
1 keV compared to $FWHM \sim 110$ eV at 1 keV 
for the {\it ASCA} SIS, during the Ton S~180 observation.  
The {\it BeppoSAX} model is also inconsistent with the 
{\it ASCA} data, primarily because the former lacks any 
parameterization of the soft spectral hump. 
As noted by Fink \etal\ (1997), the source exhibits 
spectral variations, so 
we should not expect perfect consistency between the {\it ROSAT}, {\it BeppoSAX} 
and {\it ASCA} data. 

In an attempt to gain further insight, we examined the 
extrapolation of various models  
and compared them to infrared data; optical data; 
ultraviolet data from both  the International Ultraviolet Explorer 
({\it IUE}) plus {\it EUVE} 
and PSPC data (Fig.~6). All the fluxes were 
absorption-corrected  and taken from Vennes \etal\ (1995).
It is obvious that the broken power-law model provides the 
least satisfactory extrapolation to lower energies. The steep 
part of the broken power-law is $\Gamma \sim 3.7$. 
It can be seen that for Ton S~180 such a steep component overpredicts the 
{\it EUVE} flux by an order of magnitude.  
If the soft X-ray continua are this steep, then 
there is a flattening of the spectrum within the {\it ROSAT} PSPC bandpass. 
 However, 
we reject this solution, based upon the large discrepancy of this model 
compared to the 0.4--0.6 keV  {\it ASCA} data. 
The continua from the other two models are consistent with 
a simple extrapolation down to the {\it IUE} bandpass, within a factor 
consistent with the source variability amplitude, and do not require any 
additional component peaking in the (unobserved) XUV regime 
(between $\sim 0.01$ and 0.1 keV). In these cases the model has to turn 
over close to the {\it IUE} bandpass, in order to explain the 
infrared and optical data. 
We return to these points in the discussion. 

\subsection{The Iron K$\alpha$ Line}

Analysis of the iron K$\alpha$ line is particularly interesting in the 
light of the {\it BeppoSAX} result. Comastri \etal\ (1998) found 
evidence for line emission at $\sim 7$ keV, indicative of an 
origin from material containing ionized iron, consistent with the iron 
recombination line from H-like iron at 6.94 keV.      
So, having found an adequate parameterization of the continuum shape, we 
returned the 5.3 -- 8.0 keV (rest-frame) data to the analysis, and examined the 
data/model ratio (versus the best-fit continuum model). Fig.~7
shows this ratio, which indicates the presence of significant line emission 
attributable to iron K$\alpha$, apparently peaking 
close to $\sim 7$ keV in the rest-frame, confirming the {\it BeppoSAX} result. 
We fit the line with a redshifted, 
narrow gaussian component which yielded a reduction in 
$\chi^2$ by 18, giving $\chi^2/dof=882/851$, compared to a model 
without the component. The rest-energy was 
$E=6.92^+_-0.23$keV for a line of equivalent 
width $EW=316^{+198}_{-201}$ eV.  
However, the line profile appears broad, and 
allowing the width of the model line to 
be free achieved a further reduction in $\chi^2$ by 5, significant 
at $>95\%$ confidence. That 
fit yielded a rest-energy $E=6.44^{+1.06}_{-0.44p}$keV, where 
$p$ indicates the error search hit the limit allowed for the parameter. 
The line width was large $\sigma=1.0^{+0p}_{-0.88}$keV, and 
$EW=963^{+361}_{-692}$ eV. Such a large width, and discrepancy between 
the best-fit energy using narrow versus broad-line models indicates the 
asymmetry of the line is significant. Thus we attempted  
some further models. We replaced the broad gaussian with two narrow gaussian 
lines. One had the line energy fixed at a rest-energy 6.4 keV, 
the other had the energy free. 
This fit gave no statistical improvement over the broad-line model, 
and had two more degrees of freedom. It 
yielded a rest-energy 
$E=7.01^{+0.24}_{-0.21}$ keV, and $EW=333^{+204}_{-210}$eV plus 
an equivalent width for the 6.4 keV line of $EW=150^{+156}_{-150}$ eV. 
This indicates that emission from highly ionized material provides a 
significant contribution to the overall iron K$\alpha$ flux, which 
is also evident from inspection of Fig.~7. 

The asymmetry of the profile prompted us to fit the iron K$\alpha$ line using 
the disk-line model profile as described in \S 4.1. 
This fit gave $i=39^{\rm o}\ ^{+21}_{-39}$, 
$E=6.61^{+0.25}_{-0.21}$ keV (rest-energy) and 
$EW=814^{+438}_{-869}$ eV, with a fit-statistic similar to that obtained 
with the broad gaussian parameterization, with the same number of degrees 
of freedom. We note that the inclination implied from fitting the 
soft line with this model is inconsistent with the inclination inferred 
from fitting the iron K-shell line. The most likely possibility seems to be that 
the soft X-ray feature is not dominated by a single species originating 
in an accretion disk. 

\section{Spectral Variability}

Ton S~180 shows a significantly greater $\sigma^{2}_{rms}$ in the soft band 
compared to the 
hard band, as described earlier. This indicates the spectrum of the source 
is variable, so we selected spectra from 
high and low states for comparison. 
First we split the data using an intensity division  
equivalent to an SIS0 count rate of 0.65 ct/s which yielded 
high-state and low-state spectra with a similar number of counts 
in each. 
However, no significant difference was seen in any parameter when using this
intensity selection. Next we isolated data taken during the peaks of 
the two flares, as indicated on Fig.~1, and compared those spectra with the 
remainder of the data. This  yielded ``flare'' and ``quiescent''  
spectra with 0.5--2 keV fluxes 
$1.40 \times 10^{-11} {\rm erg\ cm}^{-2} {\rm s}^{-1}$
and 
$1.08 \times 10^{-11} {\rm erg\ cm}^{-2} {\rm s}^{-1}$, respectively, 
and 2--10 keV fluxes 
$6.96 \times 10^{-11} {\rm erg\ cm}^{-2} {\rm s}^{-1}$
and 
$5.85 \times 10^{-12} {\rm erg\ cm}^{-2} {\rm s}^{-1}$, respectively. 
Thus the ``flare'' state is 20\% higher than the ``quiescent'' state in the 
2 -- 10 keV band, and  30\% higher in the 0.5 -- 2 keV band. 

No significant variability was seen in the soft-bump.
However, the spectral slope appeared slightly steeper and the Fe 
K$\alpha$ line stronger 
during the flares. These two parameters are 
very highly correlated in these fits. To determine which 
component was most likely variable 
we calculated the 90\% confidence contours  for 
photon index versus line intensity for the 
Fe K$\alpha$ line (Fig.~8). 
These contours suggest that the line is more intense during the 
flare state, while the suggested steepening of the photon index appears 
less significant. We also find that the equivalent width 
of the line appears to {\it increase} with flux. The equivalent width 
during the high-state lies above the 90\% confidence range determined for 
the low-state; we find  
$EW_{low}=421^{+700}_{-328}$ eV and $EW_{high}=1630^{+760}_{-1078}$ eV. 
Thus, while there appears to be a global X-ray Baldwin effect observed 
across samples of QSOs, whereby the 
equivalent width of the iron K$\alpha$ line is seen to reduce with increasing 
luminosity (Iwasawa \& Taniguchi 1993), this does not appear to be true for 
luminosity variations in this object and in fact the opposite behaviour 
is observed. An increase in $EW$ with increased flux indicates a 
significant change in the conditions of the reprocessor 
with changing flux. One possible explanation is 
rapid changes in the fluorescent yield of the reflector, giving more 
line photons in the high-state because the material is more highly ionized. 
If true, this would be accompanied by a shift of the line peak to higher 
energies 
as the source flux increases. Unfortunately we 
cannot determine whether such an energy shift 
occurs, using these data. 
These results on changes in the iron K$\alpha$ line should be considered 
tentative, 
as the fit parameters indicate there are only $\sim 200$ photons in the 
line during the times selected for the flares, and $\sim 350$ during 
the remainder of the observation. 

\section{Summary of Observational Results}

We present an {\it ASCA} observation of Ton S~180 which shows:
\begin{itemize}
\item Factor-of-two variability in X-ray flux over a few thousand 
seconds, with significantly 
higher $\sigma^{2}_{rms}$ in the 0.5--2 keV band, than in the 
2--10 keV band. This source also shows much higher $\sigma^{2}_{rms}$
than other Seyfert galaxies, predominantly due to the greater 
degree of variability in the soft band.  
\item A power-law with photon index $\Gamma \sim 2.5$ provides an 
adequate parameterization 
of the continuum across 0.6--10 keV with no intrinsic X-ray absorption 
\item A complex soft X-ray spectrum which can be parameterized as 
a blend of line emission peaking in the 0.7-1.0 keV regime
\item A confirmation of Fe K-shell emission from highly-ionized Fe in this 
source. Furthermore, the {\it ASCA} data show the line profile to be 
broad and asymmetric 
\item A tentative suggestion of rapid variability in the iron K-shell line, with 
an increase in line flux and equivalent width during X-ray flares. 
\end{itemize}

\section{Discussion} 

Ton S~180 shows larger amplitude variations  than
Seyfert~1 galaxies of comparable luminosity, when sampled 
on a timescale of 128 s. 
Variations of large amplitude 
could indicate that we are seeing X-rays from a more compact region 
in NLSy1 galaxies, and if we assume these X-rays originate in the accretion 
disk 
then either the disk is more compact or perhaps extends closer to 
the black hole. The latter would be possible if the hole has a greater 
degree of rotation in NLSy1 galaxies than in Seyfert~1 galaxies,
 as suggested by authors such as 
Forster \& Halpern (1996). Another simple possibility is that in NLSy1s  
the bulk of the X-ray emission is seen directly, or with fewer scatterings  
than the emission observed from  Seyfert~1 galaxies of the same luminosity. 
This could be a result of a physical difference in the scattering region  
in NLSy1 versus other Seyfert galaxies, or could be a result of 
differences in the orientation 
for the two types of Seyfert such that light paths to the observer are 
different. Instabilities can occur in the inner 
accretion disk when the accretion rate is greater than a small fraction of 
maximum rate allowed by the Eddington limit (e.g. Shakura \& Sunyaev 1976), and 
these could manifest themselves as variations in X-ray flux. 
So, another interesting possibility is that the 
differences in spectral and variability properties 
between NLSy1 and Seyfert~1 galaxies both originate as a result of a 
higher accretion rate in NLSy1s. 

Ton S~180 also shows higher variability ($\sigma^{2}_{rms}$) in the soft band, than 
the hard. This behavior is like that observed in 
NGC~4051, Mkn~766, NGC~3227 and NGC~3516 (N97a). 
In NGC~3227 this is attributed to changes in the 
emission components rather than the 
absorber (George \etal\ 1998b). In NGC~3516 we know that 
$FWHM$  H$\beta \sim 6800$ km/s (Osterbrock 1977), and 
the observation of this behaviour 
in NGC~3516 indicates the phenomenon 
is not restricted to Seyferts with narrow H$\beta$. 
For Ton S~180, the absence of any significant 
intrinsic X-ray absorption means that 
the stronger variations in the soft versus hard flux are unlikely 
to be attributable to variable opacity in an absorber.  This in turn 
indicates that the difference is likely to be due to 
changes in the continuum slope or the relative strength of emission 
components, as inferred for NGC~3227 (George \etal\ 1998b). 
A {\it possible} interpretation is that the photons at higher X-ray energies 
are produced by upscattering of soft photons, the scattering process 
could dilute the observed variability by an amount dependent of the size 
of the scattering region. 
Alternatively there may be more or larger perturbations in the region where the 
soft photons are produced. We were unable to satisfactorily 
quantify the spectral variability through flux-divided spectral analysis. 

The mean spectrum reveals a photon index $\Gamma \sim 2.5$ extending to 
10 keV. This supports the conclusion by Brandt \etal\ (1997) that 
sources with narrow permitted lines have steeper continua 
in the 2 -- 10 keV band than those with broad permitted lines. 
Furthermore, Ton S~180 is consistent with the distribution for
$\Gamma$ versus H$\beta$, shown in Brandt \etal\ (1997). 

We find strong evidence for a 
complex soft X-ray spectrum, confirming indications from 
PSPC data (Fink \etal\ 1997). 
Ton S~180 shows a hump of X-ray emission peaking at a rest-energy 
close to 0.9 keV. This could be due to a blend of line emission, 
predominantly from 
ionized species of Ne and from the Fe L-shell. The observed feature is 
broadly consistent with thermal emission from a  
plasma of temperature $\sim 0.84$ keV. 
The soft X-ray spectrum is also qualitatively similar to that 
predicted when nuclear radiation is reprocessed in small, dense clouds 
that are optically thin to Thomson scattering (Kuncic 
\etal\ 1997). 
Somewhat similar spectral complexity is also observed in quasars 
(Nandra \etal\ 1998), indicating this may be a common phenomenon 
and an important component of the X-ray spectrum. 
The observed hump could be the peak of 
emission from a hot continuum component, however, contrary to 
Nandra \etal\ 1998) we find a blackbody component to be too 
broad to satisfactorily explain the observed feature in this source. 

The emission from an ionized accretion disk yields a spectrum rich
in emission lines 
(e.g. Zycki \etal\ 1994;  Ross \& Fabian 1993; Matt, Fabian \& Ross 1993) 
for some accretion rates. 
This means that the presence of emission lines 
can be an important indicator of the accretion rate in a source, 
although unfortunately we cannot usefully constrain that quantity 
using these data, and with current models. 
For high accretion rates, the disk becomes 
highly ionized. An ionized  
disk has a reduced cross-section at low energies, so the 
contrast between the continuum and reflected emission will be greatly reduced. 
In the case of Ton S~180 the continuum is very steep, giving 
rise to relatively few photons above $\sim$10~keV from which to 
produce a Compton Reflection hump. 
The absence of a hard spectral 
component, and the profile of the iron K-shell line are 
consistent with the picture of an ionized disk in this 
source. 

We confirm the discovery by Comastri \etal\ (1998) of 
iron K-shell emission peaking close to 7 keV. Furthermore, these 
{\it ASCA} data show the iron K-shell line to be broad and asymmetric. 
The shape of the line profile suggests an origin in an accretion disk, 
rather than a blend of unresolved narrow lines from regions further out. 
While the individual parameters are poorly constrained in the diskline fit, 
the line is clearly significantly different to the mean profile obtained for 
Seyfert~1 galaxies, which could be due to either a difference in ionization or 
orientation of the reprocessor. We favor higher 
ionization of the disk as an explanation for two reasons. First, the high 
equivalent width of the line is difficult to explain if the disk is edge-on, 
as absorption of the line photons is expected within the disk (e.g. George \& Fabian 
1991). Second, the soft X-ray spectrum supports the existence of an ionized
reprocessor in this source. Whilst we did not find a good fit to the soft 
X-ray feature using a single diskline in \S4.1, 
as stated above, ionized disk models
can predict emission from several highly-ionized species in the 
$\sim$0.7--1.5~keV band.
It is therefore possible that the soft emission feature is a blend
of such lines; unfortunately the quality of the {\it ASCA}\ data is 
insufficient to allow the deconvolution of such a blend.

This result is particularly intriguing 
in the light of the apparent correlation between the profile of the 
Fe K-shell line and 2--10 keV luminosity, 
for a sample of quasars (Nandra \etal\ 1997b). Those authors find that the 
red wing and 6.4 keV peak of the line decrease with increasing nuclear 
luminosity, which is an extension of the X-ray Baldwin effect.  
The 2 -- 10 keV luminosity of Ton S~180 is  
$\sim 10^{44} {\rm erg\ s}^{-1}$ putting 
Ton S~180 on the borderline where line profile changes start to be seen in
the quasar sample (Nandra \etal\ 1997b). 
Of course some distribution is expected on a source-by-source basis,  
if not in the intrinsic relationship then certainly 
in the observed relationship, since 
there could be some difference in the luminosity which is observed 
and that which is incident on the reprocessor. 

Time-resolved spectroscopy tentatively suggests 
the strength of the iron K$\alpha$ line increases when the continuum 
flares occur. 
If confirmed, the rapid response of the line 
would strongly confirm its proposed origin in an ionized disk, 
the bulk of the line arising within a few light 
hours of the active nucleus. As noted by Comastri \etal\ (1998), 
relativistic boosting would be expected to affect the iron K-shell line 
if it is produced very close to the black hole, thus the correlation of 
line and continuum would suggest the flare events are due to perturbations 
close to the central black hole, while the longer timescale variations 
are fundamentally different in nature and are not 
necessarily accompanied by changes in the line strength. 

The complexity of the spectrum and the 
cutoff of the (reliable) {\it ASCA} data at 0.6 keV leave us insensitive to 
any continuum softening, and we do not require a steepening of the power-law 
continuum within the 0.6 -- 10 keV band. 
Laor \etal\ (1997) dispute the existence of a large bump peaking in the XUV, 
based upon {\it ROSAT} observations of quasars. 
As demonstrated in Fig.~6, our preferred solution is consistent with a simple 
extrapolation of the {\it ASCA} continuum down to the {\it IUE} bandpass, 
in agreement with the Laor result (allowing 
that the {\it EUVE} point is high by a factor which is consistent with the source
amplitude of variability). Korista, Ferland and Baldwin (1997) have
discussed the general problem that extrapolating the 
known soft X-ray spectrum of AGN down to the {\it EUVE} and {\it IUE} 
bands  
there appear to be too few 54.4 eV photons to account for the strength of 
observed He{\sc ii} lines. They consider the possibility that the broad-line
clouds see a different continuum than the observer does, or 
that the XUV spectrum
has a double-peaked shape. {\it AXAF} and {\it XMM} observations of this 
source are clearly required, to unambiguously resolve the X-ray emission lines 
and thus to allow us to tightly constrain the 
X-ray continuum in order to determine whether this component really 
does extrapolate to meet the UV data. 

\section{Acknowledgements}
We are grateful to \asca\ team for their operation of the satellite
and to the referee, Smita Mathur, for comments which significantly 
improved the paper. This research has 
made use of the NASA/IPAC Extragalactic database,
which is operated by the Jet Propulsion Laboratory, Caltech, under
contract with NASA; of the Simbad database, 
operated at CDS, Strasbourg, France; and data obtained through the 
HEASARC on-line service, provided by NASA/GSFC. We acknowledge the 
financial support 
of Universities Space Research Association (IMG, TJT) and 
the National Research Council (KN).

\newpage

\clearpage
\pagestyle{empty}
\setcounter{figure}{0}
\begin{figure}
\plotfiddle{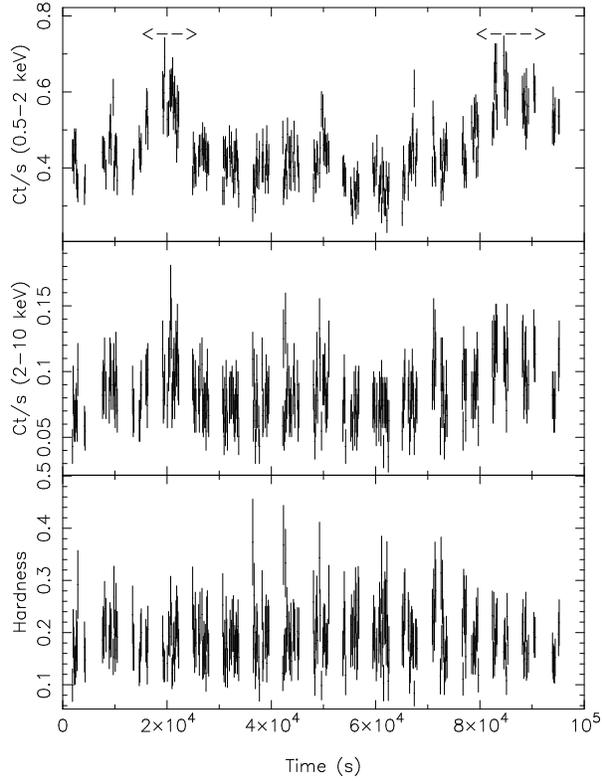}{10cm}{0}{50}{50}{-200}{0}
\caption{
\label{fig:1} 
The light curves in 128 s bins for the combined SIS data in the 
0.5-2 keV band (top panel) and 2-10 keV band (middle panel). 
The ratio of 2-10 keV/0.5-2 keV counts is shown in the panel. 
The arrows indicate the periods of data taken to be the ``flare-states'', 
as discussed in \S5.}
\end{figure} 
\clearpage

\setcounter{figure}{1}
\begin{figure}
\plotfiddle{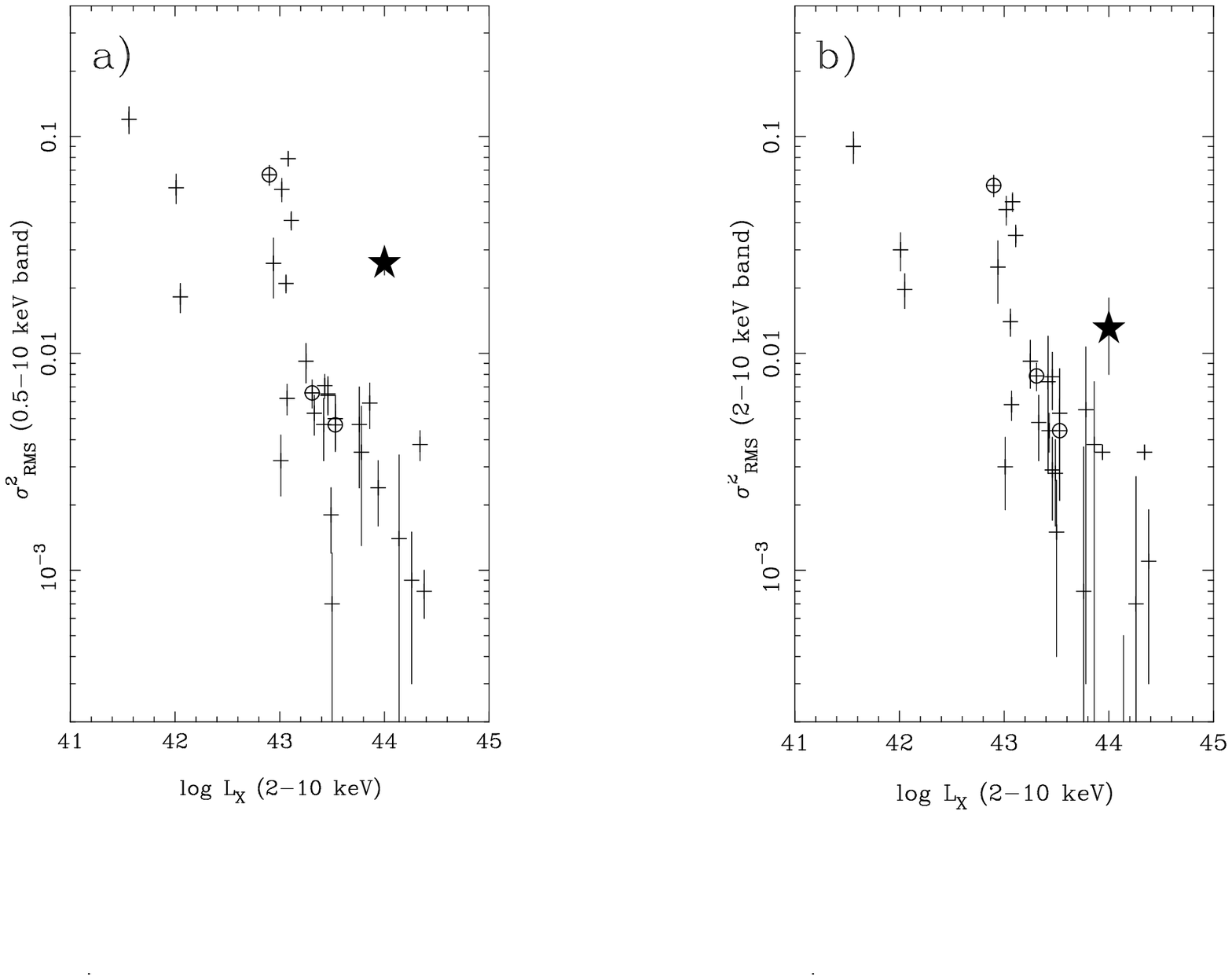}{10cm}{0}{50}{50}{-200}{0}
\caption{
\label{fig:2}
a) Normalized excess variance ($\sigma^2_{rms}$), taken over 
the 0.5-10 keV band, versus luminosity 
for the Seyfert~1 sample from N97a; 
George \etal\ 1998a,b (crosses); a few Seyfert type 
2 galaxies from Turner \etal\ 1997 (crossed circles) and 
Ton S~180 (star) which lies significantly above the rest of the
distribution; 
b) in this plot $\sigma^2_{rms}$ is taken over 
the 2-10 keV band and Ton S~180 lies close to the rest of 
the distribution.} 
\end{figure}
\clearpage

\setcounter{figure}{2}
\begin{figure}
\plotfiddle{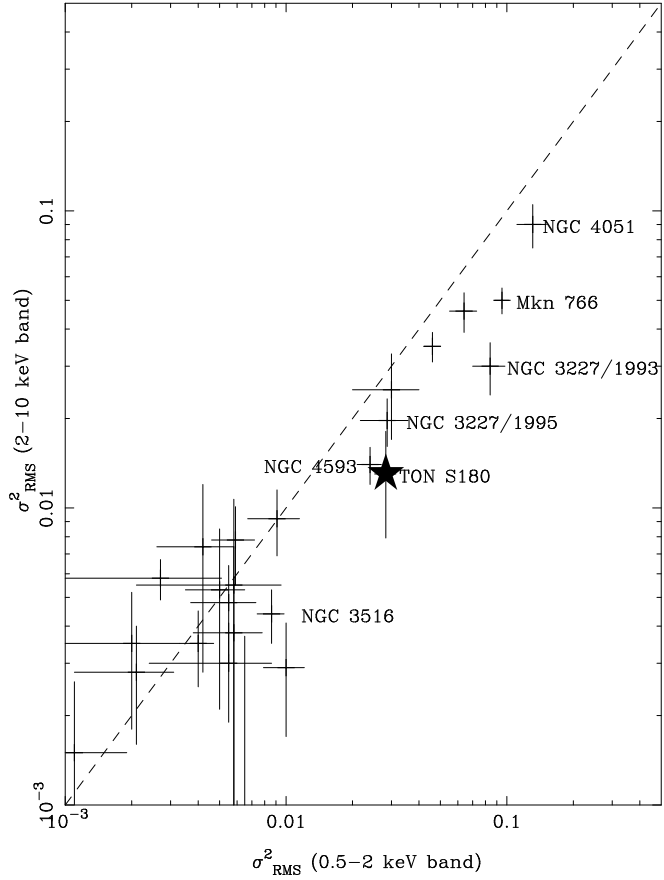}{10cm}{0}{50}{50}{-200}{0}
\caption{
\label{fig:3}
The excess variance $\sigma^2_{rms}$ in the soft band, 0.5-2 keV, versus that in the hard 
band, 2-10 keV, for the SIS data.  The Seyfert~1 sample (N97a; 
George \etal\ 1998a,b) are 
shown as crosses. Ton S~180 is shown as a star, and has a higher 
$\sigma^2_{rms}$ in the soft band than in the hard, as discussed in the text.
}
\end{figure}
\clearpage

\setcounter{figure}{3}
\begin{figure}
\plotfiddle{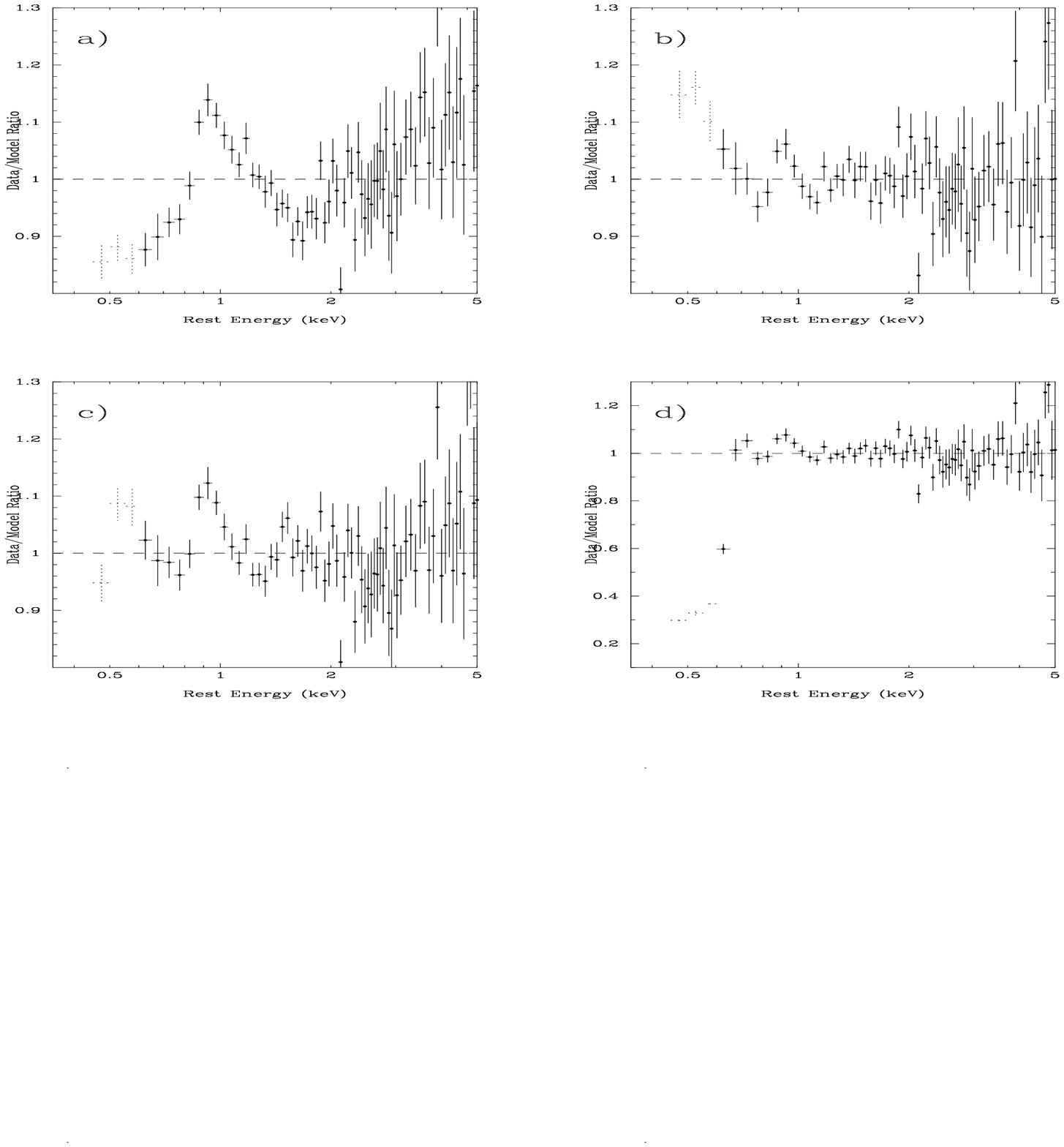}{10cm}{0}{0.5}{0.5}{-200}{0}
\caption{
\label{fig:4}
a) The data/model ratio from the combined SIS+GIS data, 
compared to a power-law model.  The dotted 
points show the SIS data from the 0.4-0.6 keV band, which were not used 
in the fit but have been overlaid for illustrative purposes; 
b) data are compared to power-law model with 
broad gaussian emission line; 
c) data are compared to a power-law plus two absorption edges; 
d) data are compared to a broken power-law model 
with a single edge, showing 
the large discrepancy between this model and the low energy data.}
\end{figure}
\clearpage

\setcounter{figure}{4}
\begin{figure}
\plotfiddle{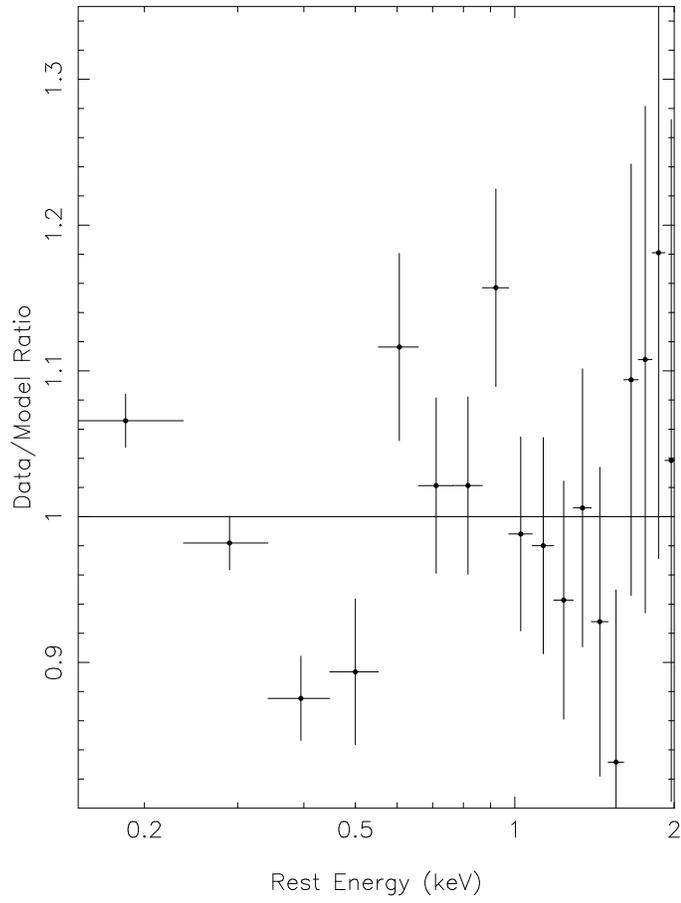}{10cm}{0}{50}{50}{-200}{0}
\caption{
\label{fig:5}
The data/model ratio compared to a power-law model, for the 
{\it ROSAT} PSPC data. 
}
\end{figure}
\clearpage

\setcounter{figure}{5}
\begin{figure}
\plotfiddle{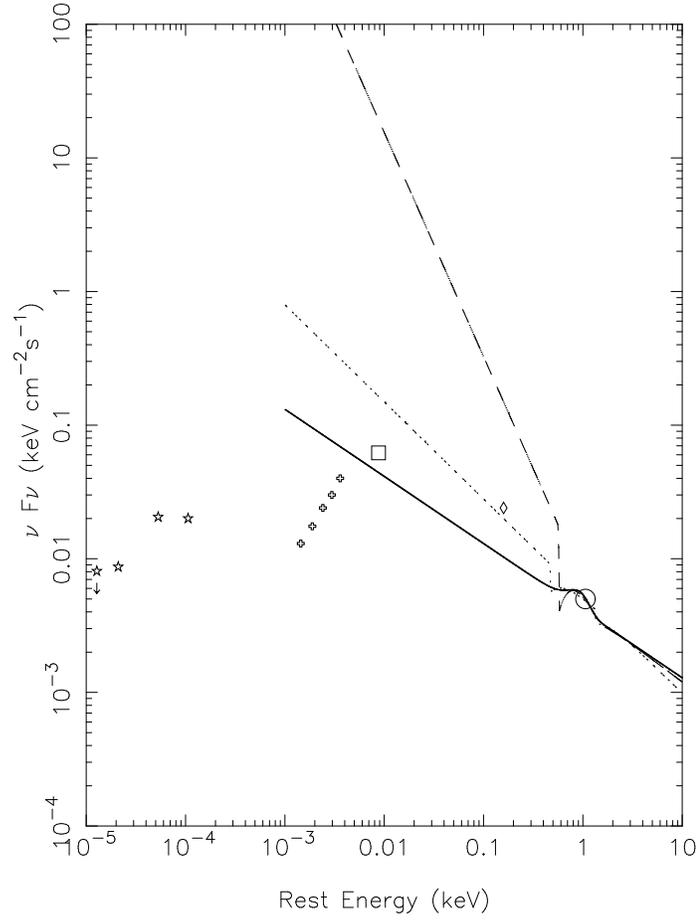}{10cm}{0}{50}{50}{-200}{0}
\caption{
\label{fig:6}
Three models for the {\it ASCA} data, 
compared to the absorption-corrected infra-red data (open stars), 
the optical U,V,B,R,I fluxes (open crosses), 
{\it IUE} data (open box), {\it EUVE} data (open diamond) and the 
{\it ROSAT} PSPC data (open circle). The dashed model line 
represents the broken power-law with single edge; the 
dotted line represents the single power-law with two edges; the 
solid line represents the single power-law plus emission feature. 
}
\end{figure}
\clearpage

\setcounter{figure}{6}
\begin{figure}
\plotfiddle{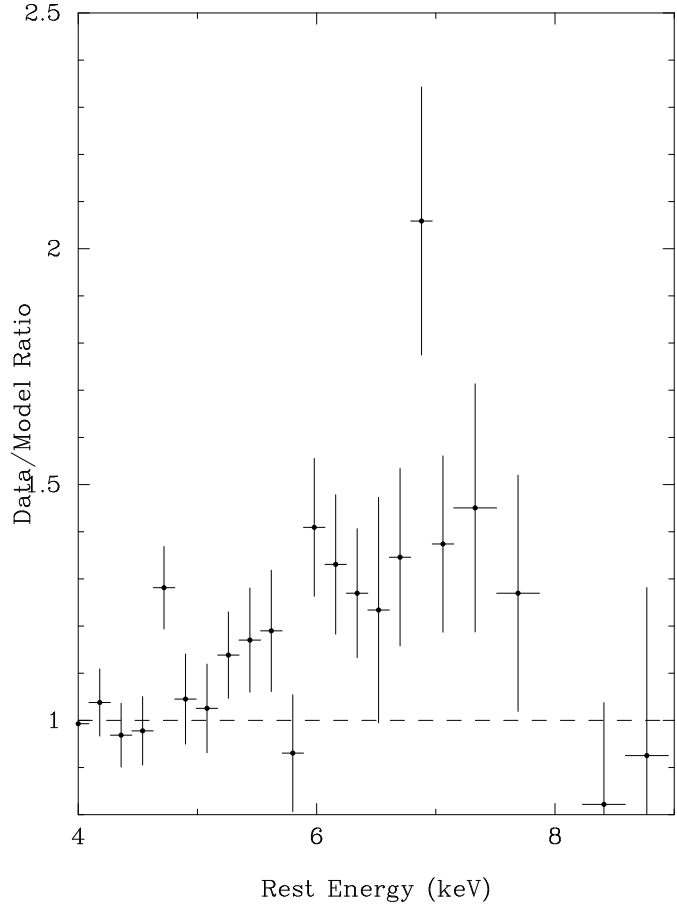}{10cm}{0}{50}{50}{-200}{0}
\caption{
\label{fig:7}
Data/model ratio based on the combined SIS+GIS data, transformed into the 
rest-frame. The 
continuum was fit without the (rest-frame) 5.3 -- 8 keV data, using the 
power-law model with the soft hump fit by a broad gaussian (as 
described in the text). The 5.3 - 8.0 keV data were then overlaid 
and have been rebinned for clarity. These data indicate line flux 
distributed asymmetrically across the 5.0 - 8.0  keV range 
with a peak close to a rest-frame energy of 7 keV.}
\end{figure}
\clearpage

\setcounter{figure}{7}
\begin{figure}
\plotfiddle{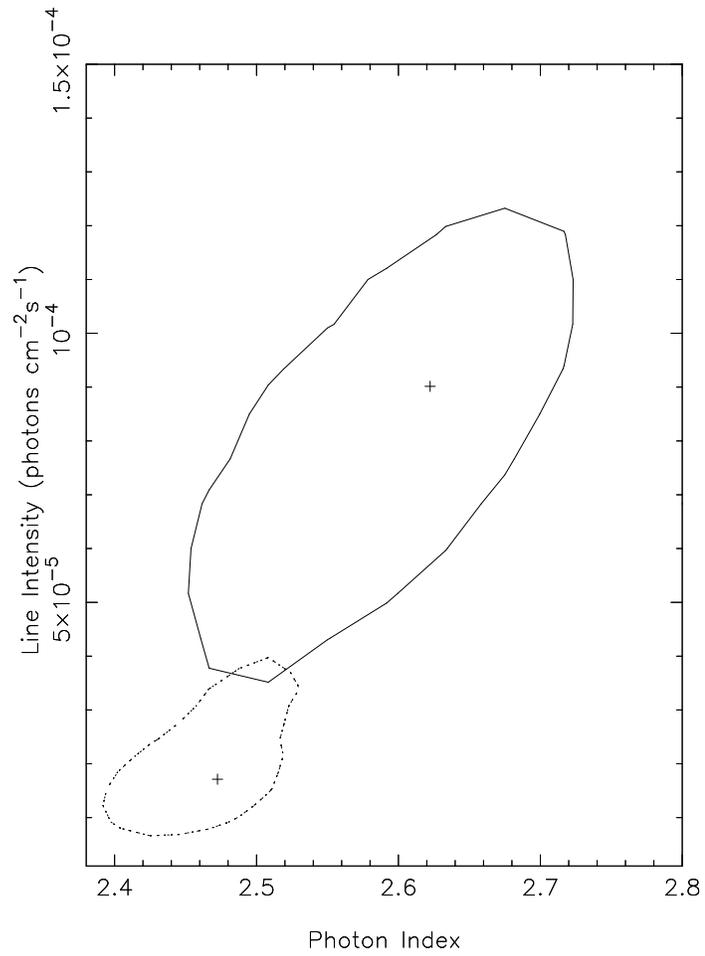}{10cm}{0}{50}{50}{-200}{0}
\caption{
\label{fig:8}
Contours indicating the 90\% confidence limit for photon index 
versus intensity of the iron K-shell line. 
The solid line represents the flare-state while 
the dotted line represents the remainder of the 
observation. The best-fitting values are indicated by crosses. }
\end{figure}


\clearpage


\begin{references}

\reference{boll} Boller, T., Brandt, W.N., Fink, H.H., 1996, A\&A, 305, 53
\reference{Brandt} Brandt, W.N., Mather, S., Elvis, M., 1997, \mnras, 285, L25
\reference{Bow96} Bowyer, S. \etal\ 1996, A\&A, 305, 53 
\reference{Bu94} Burke, B.E., Mountain, R.W., Daniels, P.J., Dolat, V.S.,
                1994, IEEE Trans. Nuc. SCI. 41, 375
\reference{com} Comastri, A. \etal\ 1998, A\&A, in press
\reference{fab} Fabian, A.C., Rees, M.J., Stella, L., White, N.E. 1989,
        \mnras\, 238, 729
\reference{fnk} Fink. H.H., Walter, R., Schartel, N., Engels, D., 
	1997, A\&A, 317, 25
\reference{fh} Forster, K., Halpern, J., 1996, \apj, 468, 565 
\reference{Ge} George, I.M., Mushotzky, R.F., Turner, T.J., Yaqoob, Y., 
	Ptak, A., Nandra, K., Netzer, H., 1998b, ApJ, submitted
\reference{Ge} George, I.M., Mushotzky, R.F., Turner, T.J., Netzer, H. 
	Nandra, K., 1998a, ApJ, in press 
\reference{Hal} Halpern, J.P. \& Oke, J.B., 1987, ApJ, 312, 91
\reference{hw} Hwang, C-Y., Bowyer, S., 1997, ApJ, 475, 552 
\reference{it} Iwasawa, K. \& Taniguchi, Y., 1993, \apjl, 413, 15
\reference{kfb} Korista, K., Ferland, G. \& Baldwin, J., 1997, ApJ 487, 555 
\reference{kcr} Kuncic, Z., Celotti, A., Rees, M., 1997, \mnras, 284, 717 
\reference{la} Laor, A., Fiore, F., Elvis, M., Wilkes, B.J., 
	McDowell, J.C., 1997, ApJ 477, 93
\reference{law} Lawrence, A., Elvis, M., Wilkes, B.J., McHardy, I., Brandt, 
	W.N., 1997, \mnras, 285, 879
\reference{lied} Liedahl, D.A., Osterheld, A.L., Goldstein, W.H., 1995, 
	ApJ, 438, L115
\reference{lei} Leighly, K., Mushotzky, R., Nandra, K., Forster, K., 1997, 
	ApJ, 489, L25 
\reference{Ma96} Makishima, et al
                1996, \pasj, 48, 171
\reference{Mt} Matt, G., Fabian, A.C., Ross, R.R., 1993, \mnras\, 264, 839 
\reference{Mew} Mewe, R., \& Kaastra, J.S., 1992, ``An X-ray Spectral Code 
	for Optically Thin Plasmas'', an internal SRON-Leiden Report, v 2.0)
\reference{n97a} Nandra, K., George, I.M., Mushotzky, R.F., Turner, T.J.,
                Yaqoob, T., 1997, \apj, 476, 70 (N97a) 
\reference{n97c} Nandra, K., George, I.M., Mushotzky, R.F., Turner, T.J., 
	Yaqoob, T., 1997, ApJ 488, L91 (N97b)
\reference{n98} Nandra, K., George, I.M., Mushotzky, R.F., Turner, T.J., 
	Yaqoob, T., 1998, in prep 
\reference{Oh96} Ohashi, T., et al.,
                1996, \pasj, 48, 157
\reference{orr} Orr, A., Yaqoob, T., Parmar, A.N., Piro, L., White, N.E., 
	Grandi, P., 1998, \aap, submitted. 
\reference{oster} Osterbrock, D.E., 1977, ApJ 215, 733 
\reference{pd95} Pounds, K.A., Done, C., Osborne, J.P., 1995, \mnras, 277, L5
\reference{r} Ross, R.R., Fabian, A.C. 1993, \mnras\, 261, 74 
\reference{ss76} Shakura, N.I., Sunyaev, R.A., 1976, \mnras, 175, 613 
\reference{stark} Stark, A.A., Gammie, C.F., Wilson, R.W., Bally, J., 
	Linke, R.A., Heiles, C., Hurwitz, M., 1992, ApJS, 79, 77
\reference{tua} Turner, T.J., George, I.M., Nandra, K. \&
        Mushotzky, R.M., 1997, \apj\ in press
\reference{Ven} Vennes, S., Polomski, E., Bowyer, S., Thorstensen, J.R., 1995, ApJ, 448, L9
\reference{wis} Wisotzki, A., Dreizler, S., Engels, D., Fink, H.H., 
	Heber, U., 1995, A\&A, 297, L55
\reference{Zy} Zycki, P.T., Krolik, J.H., Zdziarski, A.A., 
	Kallman, T.R., 1994, ApJ, 437, 597

\end{references}
\end{document}